\documentclass[aps,preprint,groupedaddress]{revtex4}
\usepackage{amsfonts}
\usepackage{amssymb}
\usepackage{graphicx}
\newcommand{\ds}{\displaystyle}
\newcommand{\qbq}{\mbox{$q\bar{q}$}}
\newcommand{\qvec}{\mbox{\boldmath $q$}}

\newcommand{\pvec}{\mbox{\boldmath $p$}}
\newcommand{\rvec}{\mbox{\boldmath $r$}}
\newcommand{\xvec}{\mbox{\boldmath $x$}}

\newcommand{\sigmavec}{\mbox{\boldmath $\sigma$}}
\newcommand{\tauvec}{\mbox{\boldmath $\tau$}}
\newcommand{\pivec}{\mbox{\boldmath $\pi$}}

\begin{document}
\title{Correlated two-pion exchange and large-$N_c$ behavior
       of nuclear forces}
\author{Murat~M.~Kaskulov}
\email{kaskulov@pit.physik.uni-tuebingen.de}
\author{Heinz~Clement}
\email{clement@pit.physik.uni-tuebingen.de}
\affiliation{Physikalisches Institut, Universit\"at  T\"ubingen,
             D-72076 T\"ubingen, Germany}
\date{\today}
\begin{abstract}
The effect of correlated scalar-isoscalar two-pion exchange (CrTPE)
 modes is considered in connection with central and spin-orbit
 parts of the $NN$ force. The two-pion correlation function is
 coupled directly to the scalar form factor of the nucleon which we
 calculate in the large-$N_c$ limit where the nucleon can be described
 as a soliton of an effective chiral theory. The results for the 
 central $NN$ force show a strong repulsive core at short internucleon
 distances supplemented by a moderate attraction beyond $1$ fm. The
 long-range tail of the central $NN$ potential is driven by 
 the pion-nucleon sigma term and consistent with the effective 
 $\sigma$ meson exchange. The spin-orbit part of the $NN$
 potential is repulsive.  The large-$N_c$ scaling behavior
 of the scalar-isoscalar $NN$ interaction is addressed. We show that
 the spin-orbit part is $\mathcal{O}(1/N^2_c)$ in strength relative
 to the central force resulting in the ratio $\simeq 1/9$ suggested by
 the $1/N_c$ expansion for $N_c=3$. The latter is in 
 agreement
 with our numerical analysis and with the Kaplan-Manohar
 large-$N_c$ power counting. Unitarization of the $\pi \pi$ scattering
 amplitude plays here an important role and improves
 the tree level results. Analytical representations of the CrTPE
 $NN$ potential in terms of elementary functions are derived
 and their chiral content is discussed.
\end{abstract}
\maketitle


\small
\section{Introduction}
  Understanding the intermediate and short range parts of the
  nucleon-nucleon ($NN$) interaction is still an interesting
  problem  and much effort is being put into this topic both from
  theoretical and from experimental sides.
  The long-range part of the $NN$ interaction is well
  described and represented by means of the one-pion-exchange (OPE)
  potential. For higher momentum transfer or small internucleon distances
  the $NN$ dynamics becomes complex and one has to rely on
  phenomenological models. In this region the  $\sigma$ meson \cite{PDG},
  for which the empirical evidence remains controversial,
  effectively represents scalar-isoscalar multi-pion
  correlations  and generates an intermediate range attraction
  in microscopic QCD motivated approaches~\cite{Jin:1996qa,PhenSigmaModels} 
  and one-boson-exchange (OBE) models~\cite{BonnPT,DeTourreil:gz}. 
  The latter  are most popular and phenomenologically
  successful in describing the nuclear force. 

  With the
  recent progress of chiral perturbation theory ($\chi$PT)
  in meson-meson and meson-baryon sectors the idea to extend
  it to the description of the $NN$ interaction appears to be quite natural.
  The chiral $SU_L(N_f) \times SU_R(N_f)$ symmetry provides a
  consistent framework for the construction of the $NN$ force 
  \cite{Weinberg:90,Ordonez:1992xp,Kaiser:1997mw}.
  In $\chi$PT the $NN$ potential has already been  calculated in
  N$^3$LO~\cite{Entem:2003ft,Higa:2003jk} and much work has been
  done in applying these ideas to nuclei, too. Supported by empirical 
  information~\cite{Rentmeester:1999vw}, the 
  $\chi$PT calculations
  are important and transparent 
  for the peripheral $NN$ partial waves,
  which are well understood in terms of OPE and two-pion-exchange (TPE)
  processes~\cite{Rentmeester:1999vw,Kaiser:1997mw,Epelbaum:2003gr}. 
  The long-range behavior of the TPE force and the effect of
  correlated $\sigma$-like scalar-isoscalar $\pi \pi$
  modes are related problems. The success
  and dynamical content of OBE models
  suggest that meson-meson correlations are 
  important~\cite{Jackson:be,Kim:1994}.
  This fact is motivated by dispersion theory, where the
  main effects of higher-order interactions can be accounted for by the
  inclusion of experimentally known meson resonances, which bring
  singularities of the amplitude close to the physical region. In
  the dispersion-theoretical framework, the $\sigma$ can be
  explained by correlated two-pion exchange with a broad spectral
  distribution around $\simeq 4 M_{\pi}$ which ultimately leads
  to the isoscalar central attraction between two nucleons.

In baryon $\chi$PT the $\pi \pi$ correlations are taken into account at tree
  level and the relation to the genuine $\sigma$ exchange is lost.
  For example, in  heavy baryon $\chi$PT
  the exchange of two uncorrelated pions, formulated using
  chiral symmetry constraints and including $\Delta$ isobars, explains
  the tail of the scalar-isoscalar $NN$ potential without any
  need for a true scalar meson \cite{Kaiser:1998wa}. Additional
  consideration of tree level $\pi \pi$ correlations leads to the 
  surprising result, that these terms
  are very small and - even more - lead to a weak 
  repulsion~\cite{Kaiser:1998wa}. Note, that in the effective theory without
  explicit $\Delta$ degrees of freedom the chiral invariant
  $NN$ interaction terms are accompanied in subleading order 
  by low-energy constants (LEC) $c_1, c_3$ and
  $c_4$~\cite{Bernard:1995dp}. In particular, the isoscalar central
  potential is dominated by 
  $c_1$ and $c_3$~\cite{Rentmeester:1999vw,Kaiser:1997mw} proportional to the
  pion-nucleon sigma term and nucleon axial-polarizability,
  respectively. The LEC $c_3$ can be approximately saturated by the
  $\Delta$ resonance. But the dimension-two
  operators $c_i$  include also information about $t$-channel meson
  exchange suggesting a strong influence of $t$-channel singularities. 
  The latter was studied in
  Refs.~\cite{Bernard:1996gq,Bernard:1993fp} 
  (see also Ref.~\cite{Bernard:1995dp}) 
  where it was shown that the 
  exchange of a scalar meson with
  mass and coupling constant similar to $\sigma$ allows to 
  explain $c_1$ and, interestingly, also part of $c_3$. Finally,
  $c_4$ mentioned above is dominated by isovector $\pi \pi$ 
  correlations from $\rho$
  meson exchange~\cite{Bernard:1996gq}. 
  Since we concentrate in this paper on $s$-wave
  CrTPE, i.e. $\sigma$- meson channel, 
  in the following we will discuss
  matters predominantly connected with $c_1$.

As well known the tree level contact
  interaction does not account for the entire $\pi \pi$ dynamics, and possibly
  one has to go beyond the tree level approximation to restore the
  relation to the $\sigma$ meson.
  Recently, non-perturbative methods describing $\pi \pi$ scattering
  were developed~\cite{Dobado:1996ps,Colangelo:2001df,
  Nieves:1998hp,Oller:1997ng,Ang:2001bd,
  He:2002ut}. One of them was proposed in Ref.~\cite{Oller:1997ti}, 
  where  based on the Bethe-Salpeter description
  of the $\pi \pi$ scattering process
  the re-summation of the infinite series of pion loops was suggested.
  This method which only uses the lowest order chiral
  Lagrangian  as input and implements an exact unitarity,
  leads to a dynamical pole in the ${\pi \pi}$ amplitude
  with a position around $M - i \Gamma/2 \simeq 450 - i 221$~MeV.
  Naturally, it is identified with the $\sigma$ meson.
  Considering the role played by the unitary $\pi \pi$ 
  correlations (or dynamical
  $\sigma$ state) in the $NN$ interaction problem, it was 
  found~\cite{Oset:2000gn} that
  the resulting scalar-isoscalar central $NN$ potential shows 
  a strong repulsion
  at internucleon distances less then $\sim 1$~fm, and a moderate
  attraction at $r > 1$~fm. 
  This feature of unitary $\pi \pi$ scalar-isoscalar correlations
  differs from the conventional Yukawa-like $\sigma$ exchange, 
  $V_{\sigma}(r) \sim - \exp({-m_{\sigma}r})/r$,
  which always results in an attraction. 
Interestingly, in
  phenomenological $NN$ interaction models like Paris~\cite{Lacombe:dr},
  Argonne V18~\cite{Wiringa:1994wb}, Nijmegen~\cite{Stoks:wp} and
  CD-Bonn~\cite{Machleidt:2000ge} the isospin independent scalar
  components are rather similar for $NN$ separations beyond
  0.5~fm~\cite{Riska:2002vn}.  But the radial dependencies differ
  considerably at short distances, ranging from attractive for
  Nijmegen and CD-Bonn (by construction) to repulsive for the more
  phenomenological V18 and Paris potentials. It also was argued that 
  the appearance of a strong repulsion in the scalar-isoscalar channel
  cannot be interpreted in terms of meson exchange~\cite{Riska:2002vn},
  a remarkable feature which only was known from the Skyrme
  model~\cite{Skyrme:vh} and now from Ref.~\cite{Oset:2000gn}.
  It is instructive, that in the Skyrme model the use of the simplest 
  product ansatz results in a repulsive central force~\cite{Nyman:py}, 
  and the missing intermediate range attraction can be produced only
  if one goes beyond the product
  approximation~\cite{Walhout:1991sb} or additionally considers the
  correlated two-pion exchange
  (CrTPE) in the scalar-isoscalar channel~\cite{Kaiser:1989ie}.

In this work the role played by the CrTPE in the $NN$ interaction
  is reconsidered for the construction of the scalar-isoscalar central
  and spin-orbit $NN$ potentials.
  Considering the tree level and unitary two-pion correlations
  in the $NN$ interaction problem, we couple the
  $\pi \pi$ correlation function to the scalar
  form factor of the nucleon in a model-independent way.
  The latter is calculated in the limit of
  a large number of colors (large-$N_c$ limit) using the Chiral
  Quark-Soliton Model ($\chi$QSM)~\cite{Diakonov:1987ty,Schweitzer:2003sb} 
  where the nucleon can be described
  as a soliton of an effective chiral theory.
  This way our results support the structure of the scalar-isoscalar
  central $NN$ force
  observed in the unitarized $\chi$PT~\cite{Oset:2000gn} and in the Skyrme
  model~\cite{Walhout:1991sb}.
  We find a strong repulsion at short
  distances and a moderate attraction at $r> 1$~fm -  even with the tree level
  $\pi \pi$ interaction.
  The tail of the central CrTPE potential is driven by the
   the pion-nucleon ($\pi N$) sigma term and
  is consistent with the effective $\sigma$ exchange in OBE models. By this,
  the effective $\sigma NN$ coupling constant can be related to the 
  $\pi N$ sigma term.
  We also show that the above feature of the central $NN$ force is general. It
  mainly relies
  on the particular functional form of the scalar form factor of the
  nucleon and is not the effect of unitarization.
  Another point of
  interest in this work is the generation of the scalar-isoscalar 
  spin-orbit ($LS$) force. Our results for this part of the $NN$ potential show
  a repulsion - a feature which differs from phenomenological $\sigma$
  exchange resulting in an attractive $LS$ interaction.
  
The large-$N_c$ behavior of CrTPE forces is
  considered and consistency with large-$N_c$ QCD analyses and the large-$N_c$
  behavior of scalar-isoscalar components of phenomenological $NN$
  interaction models is shown. Both, the central and spin-orbit parts satisfy
  large-$N_c$ QCD counting rules. We find that the
  relative strength of spin-orbit and central interactions  scales
  according to $\simeq 1/9$ in
  remarkable agreement with the Kaplan-Manohar  large-$N_c$ spin-flavor
  power counting~\cite{Kaplan:1996rk}.
  Additionally, analytical representations of central and $LS$ potentials
  in terms of elementary
  functions are derived and their chiral content is discussed.
  We show that the chiral symmetry breaking part of the interaction is small and
  that  terms non-vanishing in the chiral limit  are
  dominant and fully drive all essential features of the
  CrTPE force in the scalar-isoscalar channel.

The paper is organized as follows. In the next two sections we
 discuss the construction of the CrTPE 
 scalar-isoscalar $NN$ force. We emphasize the importance of the
 scalar form factor of the nucleon and give its phenomenological
 description in Sec.~V. The large-$N_c$ scaling and consistency 
 are addressed in Sec.~IV. Finally,
 the analytical results and discussions are presented in Sec.~VI.

\section{Two-pion correlation function}
We start our considerations of the scalar-isoscalar CrTPE $NN$
 interaction  from the standard definition of the tree level $\pi
 \pi$ correlation function and its large-$N_c$ behavior.
 To lowest order in the derivative expansion the mesonic
 Lagrangian~$\mathcal{L}_{\pi}^{(2)}$ is given by the $SU(3)$ 
 nonlinear $\sigma$ model
\begin{equation}
\label{Lpipi}
\mathcal{L}_{\pi}^{(2)} = \frac{f_\pi^2}{4}
    \mbox{Tr} \Big[\partial^{\mu}U^{\dagger} \partial_{\mu}U \Big]
    + \mathcal{B}~ \frac{f_\pi^2}{2}
    \mbox{Tr} \Big[\mathcal{M} (U + U^{\dagger}) \Big]\ \
\end{equation}
and contains the most general low-energy interactions of the
 pseudo-scalar meson octet. In Eq.~(\ref{Lpipi})
 the leading symmetry-breaking term is linear in the quark masses
 $\mathcal{M}$ and is
 characterized by $\mathcal{B}=-\langle\qbq\rangle/f_\pi^2$. The pion-decay
 constant~$f_\pi = 93$~MeV and the scalar condensate~$\langle\qbq\rangle$ are the
 relevant parameters.  In the $SU(2)$ limit the
 non-linear field~$U$ entering Eq.~(\ref{Lpipi})
 is given by the standard matrix form:
\begin{equation}
\label{U}
U(x) = \exp \Big( i \Phi(x)/{f_\pi} \Big), ~
\Phi \equiv {\tauvec}{\pivec}
\end{equation}
where $\tauvec$ are the $SU(2)$ Pauli matrices and $\pivec$ is the 
 isovector pion field.
For the $\pi_a  \pi_b \to \pi_c  \pi_d$
scattering process, defined by the Cartesian isospin indices $a,...$, 
the use of the standard $\chi$PT procedure in expanding the
$\mathcal{L}_{\pi}^{(2)}$ to  order $\mathcal{O}(\pivec^4)$
results in the tree level contact interaction 
\begin{eqnarray}
\label{TreePiPi}
-i V_{\pi \pi}^{ab \to cd} =
\delta_{ab} \delta_{cd}
A(s)
+ \delta_{ac} \delta_{bd}
A(t)
+ \delta_{ad} \delta_{bc}
A(u),
\end{eqnarray}
where
\begin{equation}
A(s) = \frac{i}{f_{\pi}^2} \Big( s- M_{\pi}^2 - \frac{1}{3}
  \sum_{i=a,b,c,d} \Lambda_i \Big) + \mathcal{O}(q^4),
\end{equation}
and $\Lambda_{i} = k_i^2 - M_{\pi}^2$ are the off-shell part of
the invariant $\pi \pi$ amplitude. The Mandelstam variables are
related by $s + t + u = k_a^2 + k_b^2 + k_c^2 + k_d^2$. At this
order of the pion field expansion the isoscalar $S$-wave $\pi \pi$
 partial amplitude ($L=0$) is obtained from the standard decomposition
\begin{equation}
V_{\pi \pi}^{L,I=0} = \frac{1}{2} \frac{1}{(\sqrt{2})^{\alpha}}
\int_{-1}^{1} d \cos\theta~ {P}_L(\cos\theta)~ V_{\pi
\pi}^{I=0}(\theta)
\end{equation}
where ${P}_L(\cos\theta)$ are the Legendre polynomials and
$(\sqrt{2})^{\alpha}$ accounts for the
statistical factor occurring in states with identical particles:
$\alpha =2$ for $\pi \pi \to \pi \pi$. The tree level scalar-isoscalar
$\pi \pi$ scattering amplitude  is
\begin{equation}
\label{VPiPi00}
V_{\pi \pi}^{L=I=0} = - \frac{1}{f_{\pi}^2}
\Big(s - \frac{M_{\pi}^2}{2} - \frac{1}{3} \sum_i \Lambda_i \Big).
\end{equation}

In the limit of a large number of colors $N_c$ of QCD,  the scaling
behavior of Eq.~(\ref{VPiPi00}) can be obtained from large-$N_c$
$\chi$PT~\cite{Kaiser:2000gs}. The conventional large-$N_c$
counting rules require that the pion mass is $M_{\pi} \sim
\mathcal{O}(1)$ and the decay constant is $f_{\pi} \sim
\mathcal{O}(\sqrt{N_c})$. As a result the $\pi \pi$
amplitude scales according to 
\begin{equation}
\label{VPiPINc}
V_{\pi \pi} \sim \mathcal{O}(1/N_c)
\end{equation}
and large-$N_c$ QCD becomes  a theory of weakly interacting pions.
Note that the unitarization of the $\pi \pi$ correlation function
(scattering amplitude) $V_{\pi
  \pi}^{L=I=0}$, Eq.~(\ref{VPiPi00}), is a lengthy  procedure.
Here, we only mention some recent analyses which employ
dispersion relations~\cite{He:2002ut}, 
Pad\'e approximants~\cite{Ang:2001bd}, solution of
Roy~\cite{Colangelo:2001df} and
Bethe-Salpeter~\cite{Nieves:1998hp,Oller:1997ti} equations. Unitary
 $\pi \pi$ scattering amplitudes as  provided by most of these
  methods contain a dynamical pole~\cite{Dobado:1996ps} which is
  identified with the $\sigma$ meson.

\section{Two-pion exchange kernel and scalar form factor}
\begin{figure}[t]
\includegraphics[clip=true,width=0.17\linewidth,angle=90]{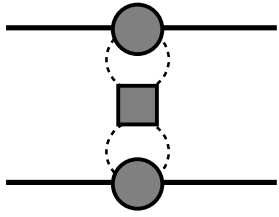}
\caption{\label{CrTPEDiagr}
\footnotesize 
The coupling of the scalar-isoscalar $\pi \pi$ correlation function
$V_{\pi \pi}^{L=I=0} $ (filled box) to the isospin symmetric part
$F^+_{\pi N}$ (filled disk) of the $\pi N$ scattering amplitude.}
\end{figure}

With Eq.~(\ref{VPiPi00}) we come to the definition of the
CrTPE potential in the scalar-isoscalar channel describing the
on-shell $NN$ scattering process, $N(p_1,s_1) N(p_2,s_2) \to
N(p_3,s_3) N(p_4,s_4)$, where $p_i$ and $s_i$ are the four momenta
and spins of interacting nucleons, respectively. As well known, in
the generic case of TPE the chiral symmetry constraints work best
and allow to relate the TPE process containing two intermediate
pions, with Cartesian isospin indices $a$ and $b$ 
to the off-mass shell $\pi N$ scattering amplitude 
in a model-independent way~\cite{Kaiser:1997mw,Higa:2003jk,Robilotta:1996ji}.
The latter is defined by the 
Green function~\cite{Gasser:1987rb}:
\begin{eqnarray}
\mathcal{G}_{\pi N} &=&
i \int \int d^4 x ~d^4 y ~\langle p',s' | \hat{T} {P}^{a}(x) {P}^{b}(y)
| p,s \rangle e^{i(k'x-ky)} \nonumber \\
&&= 
\frac{ G_{\pi}}{k^2 - M_{\pi}^2}~  \frac{ G_{\pi}}
{k'^2 - M_{\pi}^2}~ T^{ab}_{\pi
  N} (p',k',p,k) 
\end{eqnarray} 
Here $G_{\pi}$ accounts for the coupling of the pion field to
the pseudoscalar quark densities $\langle 0 | P^a(0) | \pi^b
\rangle = \delta_{ab} G_{\pi}$,
${P}^a(x)= i \bar{\psi}(x) \gamma_5 \tau^a \psi(x)$ and
$p',p~ (k',k)$ are the momenta of outgoing and incoming 
nucleons (pions). For the $\pi N$ scattering amplitude 
$T^{ab}_{\pi N}$ the standard decomposition
involves four Lorentz invariant functions $D^{\pm}, B^{\pm}:$
\begin{equation}
\label{PiNAmpl} T^{ab}_{\pi N} = \delta_{ab}~ F^{+}_{\pi N} +
\frac{1}{2} [\tau_a, \tau_b]~ F^{-}_{\pi N},
\end{equation}
\begin{equation}
F^{\pm}_{\pi N} = \bar{u}(p'_N,s') \Big[ D^{\pm} 
+ i \sigma^{\mu \nu}
  k'_{\mu} k_{\nu} B^{\pm} \Big] u(p_N,s).
\end{equation}
We are interested in the chiral content of the TPE potential in
the scalar-isoscalar channel where the introduction of the tree
level contact $\pi \pi$ interaction results in a $NN$ scattering
amplitude with two connected pion loops
\begin{eqnarray}
\label{VLI0} 
-i \mathcal{V}_{NN}
&=& i  \left(\frac{3}{2}\right) \int \int \frac{d^4 k}{(2 \pi)^4} 
~\frac{d^4
  \tilde{k}}{(2 \pi)^4}~  
\Big( V_{\pi \pi}^{L=I=0} \Big) \hspace{0.8cm}
  \\
&&\times \left[ \frac{[F^{+}_{\pi N}]^{(1)}}{(k^2 - M_{\pi}^2)(k'^2 -
M_{\pi}^2)} 
\right]
\left[ \frac{[F^{+}_{\pi N}]^{(2)}}{(\tilde{k}^2 -
M_{\pi}^2)(\tilde{k}'^2 - M_{\pi}^2)} \right]
  \nonumber
\end{eqnarray}
where $k(\tilde{k})$ and $k'(\tilde{k}')=k(\tilde{k})+p'-p$ are
the momenta of the exchanged pions, $p(p')$ are relative $NN$
momenta in the initial (final) state and the superscripts refer to
nucleons (1) and (2), respectively. $F^{+}_{\pi N}$  is the
isospin symmetric part of the $\pi N$ amplitude,
Eq.~(\ref{PiNAmpl}), and the scalar-isoscalar tree level contact
interaction $V_{\pi \pi}^{L=I=0}$ is given by the
Eq.~(\ref{VPiPi00}). The diagrammatic representation of
Eq.~(\ref{VLI0}) is shown in Fig.~\ref{CrTPEDiagr}.

The above expression, Eq.~(\ref{VLI0}) has well known
difficulties. It involves the off-mass-shell $\pi \pi$ interaction
which is not unique and depends on the choice of the interpolating
pion field (parameterization of the $SU(2)$ matrix $U$)
\cite{Kaiser:1998wa,Scherer:2002tk}. If the initial and final
pions are all on the mass shell, i.e. $\Lambda_i=0$ in
Eq.~(\ref{VPiPi00}), the $\pi \pi$ scattering amplitude is
independent of a change of field variables in agreement with the
equivalence theorem \cite{Kamefuchi:sb}. In other words
Eq.~(\ref{VPiPi00}) gives a unique result independent of the
parameterization of $U$ only for the on-shell matrix elements.
To avoid these ambiguities it was stated in Ref.~\cite{Kaiser:1998wa}
that one has to include a subset of diagrams to find cancellations
of the off-shell $\pi \pi$ isoscalar amplitude. 
This statement was rigorously verified in Ref.~\cite{Oset:2000gn},
where it was shown that the additional consideration of a subset
of chiral diagrams results in exact cancellations of the off-shell part
of the $\pi \pi$ amplitude and the on-shell residue can be
factorized out from the loop integrals of Eq.~(\ref{VLI0}).
By this and using notations of Ref.~\cite{Maekawa:2000sb} 
the $NN$ amplitude in the
general case of $L=I=0$ $\pi \pi$ exchange takes the form
\begin{equation}
\label{VNNGamma} 
\mathcal{V}_{NN}(t) =
6~[\Gamma_N^{+}(t)]^{(1)}
    [\Gamma_N^{+}(t)]^{(2)}~
\tilde{V}_{\pi \pi}^{L=I=0}(t),
\end{equation}
where $\tilde{V}_{\pi \pi}^{L=I=0}$ is the on-mass shell part of
Eq.~(\ref{VPiPi00}) with $s \to t = (p'-p)^2 = -\qvec^2$ in the
$NN$ c.m. frame. In Eq.~(\ref{VNNGamma}) the "vertex functions"
$[\Gamma_N^{+}(t)]$ are defined by
\begin{equation}
-i [\Gamma_N^{+}(t)]^{(i)} = - \frac{1}{2}
\int \frac{d^4 k}{(2 \pi)^4} 
\frac{[F^{+}_{\pi N}]^{(i)}}{(k^2 - M_{\pi}^2)(k'^2 - M_{\pi}^2)}
\end{equation}
and can be interpreted in the heavy and relativistic baryon $\chi$PT
as one-pion-loop contributions to the scalar
form factor of the nucleon $\sigma(t)$~\cite{Kaiser:1998wa,Robilotta:2000py}:
\begin{equation}
\label{GammaVertex}
[\Gamma_N^{+}(t)]^{(i)} =  \frac{1}{3}  \frac{\sigma(t)}{M_{\pi}^2}
 \left[ \bar{u} u  \right]^{(i)}.
\end{equation}
At $t=0$ the $\sigma(0)$ is referred to as the pion-nucleon
sigma-term, $\sigma_{\pi N}$. Note, that by construction the
factor $\sim 1/M_{\pi}^2$ entering Eq.~(\ref{GammaVertex}) is a
direct reflection of the chiral symmetry breaking part of the
$\chi$PT Lagrangian, Eq.~(\ref{Lpipi}). The resulting $NN$
scattering amplitude can be summarized as follows
\begin{equation}
\label{VNN} 
\mathcal{V}_{NN}(t) 
= \frac{2}{3}
\left(\frac{\sigma(t)}{M_{\pi}^2}\right)^2 \left[\bar{u} u
\right]^{(1)} \left[\bar{u} u \right]^{(2)} \tilde{V}_{\pi
\pi}^{L=I=0}(t)
\end{equation}

The quasi-potential reduction of the $NN$ amplitude
Eq.~(\ref{VNN}) is a standard procedure. The Partovi-Lomon method
is the most popular one and allows to reduce the Bethe-Salpeter
kernel to the quasi-potential containing minimal
relativity~\cite{Partovi:1997}. The latter is ready for the
iteration in the Lippmann-Schwinger equation. Here we follow
Ref.~\cite{Oset:2000gn} and wish to discuss the non-relativistic
potential only. With the normalization for Dirac spinors,
$\bar{u}(p,s) u(p,s) =1$, the non-relativistic CrTPE potential is
obtained from Eq.~(\ref{VNN}) simply by keeping its energy
independent parts:
\begin{equation}
\label{TNNNonRel}
\mathcal{V}_{NN}(t) =
\frac{2}{3} \left(\frac{\sigma(t)}{M_{\pi}^2}\right)^2
\tilde{V}_{\pi \pi}^{L=I=0}(t) \Big[ 1 - \frac{\Omega_{LS}}{2
M_N^2} \Big].
\end{equation}
In square brackets the first term gives the central  CrTPE potential
$\mathcal{V}_C(t)$, with the structure similar to that derived
in Ref.~\cite{Kaiser:1998wa}, and the second term provides the 
additional spin-orbit
part $\mathcal{V}_{LS}(t)$ with the spin content:
$ \Omega_{LS} =i (\sigmavec^{(1)}+\sigmavec^{(2)})
(\pvec' \times \pvec)/2$.
In coordinate space the central potential is given by the
Fourier transform of $\mathcal{V}_C(t)$ and the spin-orbit $(LS)$
part of the $NN$ force is related to the central potential simply
by~\cite{Partovi:1997}
\begin{eqnarray} \label{VLSCOORD} \mathcal{V}_{LS}(r) =
-\frac{1}{2 M_N^2} \frac{1}{r} \frac{\partial 
\mathcal{V}_{C}(r)}{\partial r},
\end{eqnarray}
where the spin-orbit operator $\hat{\Omega}_{LS}$ in  $r$-space
representation has been omitted. The central part of
Eq.~(\ref{TNNNonRel}) has an interesting feature. Its
long-range tail or the value of the potential at $t=0$ is
determined by the pion-nucleon sigma term: 
$\mathcal{V}_C(0)=\sigma_{\pi N}^2/3 M_{\pi}^2 f_{\pi}^2$. 
So, the only unknown element which
enters Eq.~(\ref{TNNNonRel}) is the scalar form factor of the
nucleon. As noted in Refs.~\cite{Robilotta:2000py,Maekawa:2000sb}
Eq.~(\ref{GammaVertex}) and therefore also Eqs.~(\ref{VNN})
and~(\ref{TNNNonRel}) are general and should be independent of
models or approximation schemes used to calculate the scalar form
factor of the nucleon. In this work, we calculate
$\sigma(t)$ in the large-$N_c$ limit in the framework of the 
$\chi$QSM. But before doing model
calculations the stated large-$N_c$ behavior of Eq.~(\ref{TNNNonRel})
must be understood
and consistency with large-$N_c$ QCD analyses must be shown.

\section{Large-$N_c$  scaling and consistency}
In this section we explore qualitative features of the
scalar-isoscalar components of the CrTPE force Eq.~(\ref{TNNNonRel})
which may be understood directly from
QCD. In this context, it may be useful to consider the interaction
in the limit, when the number of colors, $N_c$, of QCD becomes
large~\cite{'tHooft:1973jz,Witten:1979kh}, and to treat $1/N_c$ as
an expansion parameter. Some features of the $NN$ force can be
obtained in this limit in a model-independent
way~\cite{Kaplan:1995yg,Kaplan:1996rk,Cohen:2002qn}. Recently, it
was realized that large-$N_c$ nuclear interactions are spin-flavor
$SU(4)$ symmetric, with Wigner's supermultiplet symmetry following
as an accidental symmetry~\cite{Kaplan:1995yg}. If we assume, in
addition, that the spin-flavor symmetry properties of the $NN$
interaction are independent of phases of the many body ground
state (at $N_c \to \infty$ nuclear matter forms a classical
crystal, and a phase transition between $N_c=3$ and $N_c=\infty$
is expected~\cite{Kaplan:1996rk}), then the large-$N_c$ scaling
behavior of the $NN$ force can be analyzed in a general way
resulting in QCD expectations which can explain general features
of the $NN$ interaction~\cite{Kaplan:1995yg,Kaplan:1996rk}.

The first treatment of the $NN$ interaction in large-$N_c$ QCD was
done in Ref.~\cite{Witten:1979kh}, where it was argued that the
dominant interaction between two baryons is of order $\sim
\mathcal{O}(N_c)$. This expectation is consistent with nuclear
dynamics where the dominant interaction components
are isoscalar-scalar and -vector interactions
\begin{equation}
\mathcal{V}_{\sigma} \sim \mathcal{O}(N_c),~~~ 
\mathcal{V}_{\omega} \sim \mathcal{O}(N_c).
\end{equation}
In phenomenological models, the latter can be parameterized in
terms of effective $\sigma$ and $\omega$
exchanges~\cite{Riska:2002vn}. The entire analysis of the $NN$
potential was done in Ref.~\cite{Kaplan:1996rk} where it was found
that large-$N_c$ QCD  implies that the scalar-isoscalar central
and spin-orbit  forces scale as
\begin{equation}
\label{KMscaling}
\mathcal{V}_C \sim \mathcal{O}(N_c),~~~ \mathcal{V}_{LS} \sim \mathcal{O}(1/N_c).
\end{equation}
Recently, the dynamical interpretation of the $NN$ interaction for
four modern phenomenologically successful
models~\cite{Lacombe:dr,Wiringa:1994wb,Stoks:wp,Machleidt:2000ge}
in the large-$N_c$ limit was reported  and consistency of their
scalar-isoscalar components with Eq.~(\ref{KMscaling}) was
found~\cite{Riska:2002vn}.

In our case, the consistency condition implies that the
large-$N_c$ scaling behavior of Eq.~(\ref{TNNNonRel}) must be
consistent with large-$N_c$ QCD analysis~\cite{Kaplan:1996rk},
with the large-$N_c$ scaling of scalar-isoscalar components of all
known phenomenologically successful $NN$ interaction
models~\cite{Riska:2002vn} and also with effective $\sigma$ meson
exchange, $\mathcal{V}_{\sigma} \sim \mathcal{O}(N_c)$. 
The dynamical quantities which enter the scalar-isoscalar
CrTPE force, Eq.~(\ref{TNNNonRel}),  are the
scalar form factor $\sigma(t)$, the pion mass $M_{\pi}$, the
nucleon mass $M_N$ and the $\pi \pi$ scattering amplitude $V_{\pi
\pi}$ which scales as $V_{\pi \pi} \sim \mathcal{O}(1/N_c) $,
Eq~(\ref{VPiPINc}). We follow standard $N_c$ counting
rules~\cite{Oh:1999yj,Kaiser:2000gs} where
$\sigma(t) \sim \mathcal{O}(N_c)$, $M_{\pi} \sim \mathcal{O}(1)$,
$M_{N} \sim \mathcal{O}(N_c)$ and the nucleon momenta $|\pvec|,
|\pvec'| \sim \mathcal{O}(1)$. Counting powers of $N_c$ the central
part of the CrTPE potential, first term in Eq.~(\ref{TNNNonRel}),
is $\mathcal{V}_{C} \sim \mathcal{O}(N_c)$. Its spin-orbit
counterpart $\mathcal{V}_{LS}(t)$ additionally contains  two 
inverse powers of nucleon
mass $\sim 1/M_N^2 \sim 1/N_c^2$ resulting in $\mathcal{V}_{LS}
\sim \mathcal{O}(1/N_c)$. By this, the consistency with scaling
relations from large-$N_c$ QCD~Eq.~(\ref{KMscaling}) 
is noted. Interestingly,
Eq.~(\ref{KMscaling}) implies that the spin-orbit force
$\mathcal{V}_{LS}$ is $\mathcal{O}(1/N^2_c)$ in strength relative
to the central force $\mathcal{V}_{C}$~\cite{Kaplan:1996rk}
\begin{equation}
\label{VLStoVSscaling} \mathcal{V}_{LS}/\mathcal{V}_{C} \sim
\mathcal{O}(1/N^2_c)
\end{equation}
We refer to Eq.~(\ref{VLStoVSscaling}) as the Kaplan-Manohar
scaling relation~\cite{Kaplan:1996rk}. In the last section it will
be shown that Eq.~(\ref{VLStoVSscaling}) is very accurate
 and in remarkable agreement with our
results, 
signaling that, indeed, with
the actual number of colors $N_c=3$ the relative strength of
potentials scales like $\mathcal{V}_{LS}/\mathcal{V}_{C} \simeq
1/9$.

\section{Scalar form factor in the Large-$N_c$ limit and 
quark-soliton model}
In the large-$N_c$ limit, the QCD is equivalent to an effective
theory of mesons with baryons emerging as solitonic
configurations~\cite{Witten:1979kh}. The $\chi$QSM~\cite{Diakonov:1987ty}
provides a practical realization of the large-$N_c$ picture of the
nucleon and is considered as a chiral
relativistic quantum field
theory of quarks, anti-quarks and Goldstone bosons. It is defined
by the partition function in Euclidian space which is the
path-integral over pseudoscalar meson and quark
fields~\cite{Diakonov:1985eg,Diakonov:tw}
\begin{equation}
\label{SMPartFunc}
\label{Z}
\mathcal{Z} = \int \mathcal{D} \psi \mathcal{D} \bar{\psi}
\mathcal{D}U \exp \left[ i \int d^4 x \bar{\psi} (i \not{\partial} -
  M_qU^{\gamma_5} - m_q)
\psi \right]
\end{equation}
where $U$ denotes the $SU(2)$ pion field, Eq.~(\ref{U}), 
\begin{equation}
U^{\gamma_5} = \exp (i \gamma_5 \Phi /f_{\pi}) = \frac{1}{2} (U +
U^{\dagger}) + \frac{1}{2} (U-U^{\dagger}) \gamma_5
\end{equation}
In Eq.~(\ref{SMPartFunc}) $M_q$ and $m_q$ are dynamical, arising
from spontaneous breaking of chiral symmetry and current
quark masses, respectively.
The action, Eq.~(\ref{SMPartFunc}), was derived from the instanton
vacuum model~\cite{Diakonov:1985eg}, where the cut-off is determined
by the average size of
instantons $\langle \rho \rangle$, and the dynamical quark mass
$M_q$ is momentum dependent. In the large-$N_c$ limit the chiral
bosonic fields $U^{\gamma_5}$ can be integrated out by the saddle
point method using a
classical background field of {\it hedgehog} shape
\begin{equation}
\pivec(\xvec) = f_{\pi} \hat{\rvec}
\mathcal{P}(r)
\end{equation}
which plays the role of a Hartree-type mean
field for quarks forming a soliton-like bound state.
With $r=|\xvec|$ and $\hat{\rvec} =\xvec/r$, the variational procedure reduces
to the determination of the self-consistent soliton profile
$\mathcal{P}(r)$,
where $\mathcal{P}(r \to \infty)=0$ and $\mathcal{P}(0)=-\pi$
produce a soliton with unit winding number. In practical calculations
the soliton size $R_s=1/M_q$ is treated as a free
parameter, playing the role of the 
axial coupling constant $g_A$~\cite{Diakonov:1987ty} 
\begin{equation}
\label{GACoupl}
\lim_{r \to \infty} \left[ \lim_{m_{\pi} \to 0} r^2 \mathcal{P}(r) 
  \right] = - 2 R_s^2 = - \frac{3}{8 \pi} \frac{g_A}{f^2_{\pi}}
\end{equation}
Nucleon states of definite spin and isospin are obtained by quantizing
the rotational zero modes of the soliton.

The model expression for the scalar form factor of the nucleon in the
$\chi$QSM is quadratically ultraviolet (UV) divergent and requires
regularization~\cite{Wakamatsu:1992wr,Kim:1995hu}
\begin{equation}
\sigma(t) = m_q N_c \int d^3 \xvec j_0(\sqrt{-t} |\xvec|) \sum_{n}
\mathcal{D}_n(\xvec) \Big|_{\mbox{reg}}
\end{equation}
Here $\mathcal{D}_n(\xvec) = \Phi^{*}_n(\xvec) \gamma_0 \Phi_n(\xvec)$
is the scalar-quark density in the nucleon
and $j_0$ is the spherical Bessel function. The completeness of quark Fock
states $|\, n \rangle$ in the external pion field require
that $\mathcal{D}_n$ is represented as a sum over occupied
(valence) and non-occupied negative energy Dirac-sea states.
So it is useful to consider the contributions from discrete levels
$(lev)$ and from the Dirac-sea $(sea)$ continuum separately.
$\sigma_{lev}(t)$ is finite and requires the single-particle wave
functions $|\Phi_{lev}(\xvec)\rangle$ to be found from the Dirac 
equation in the external pion
field. The latter can be solved numerically.
$\sigma_{sea}(t)$ is UV divergent and depends on the
regularization scheme employed to make it finite. To obtain
$\sigma_{sea}(t)$ 
we follow the method
developed in Ref.~\cite{Schweitzer:2003sb}. In short, the
procedure is as follows.

To evaluate $\sigma_{sea}(t)$, the model expression for the
continuum contribution  is expanded in a series  of the
$U$-field gradients - the interpolation formula method~\cite{Diakonov:1996sr}.
The series in $\nabla U$
contains UV-divergent and UV-finite parts. As was shown in
Ref.~\cite{Schweitzer:2003sb} the latter is 
strongly suppressed with respect to the
UV-divergent term by the
instanton packing fraction or parametric smallness of 
$M^2_q \langle\rho\rangle^2 \ll
1$, and can be neglected. The dynamical quark mass is momentum 
independent and its value
$M_q=350$~MeV is taken at zero momentum transfer
from instanton phenomenology~\cite{Diakonov:1995ea}.
We recall that the $\chi$QSM is defined with some appropriate regularization.
In Ref.~\cite{Schweitzer:2003sb}  the regularization procedure 
is solved in a model
independent way by observing that the structure of divergencies in
$\sigma_{sea}$ are the same as in model expressions for the 
vacuum scalar-quark condensate
$\langle\bar{\psi}\psi\rangle$ and pion decay constant $f_{\pi}$. The
latter are fixed by their empirical values. The 
``arctan-profile'' function which provides an accurate representation of
the self-consistent profile is used 
\begin{equation}
\label{ArctPrMpi}
\mathcal{P}(r)
= -2 \arctan \left[
\frac{R^2_{s}}{r^2}~ ( 1 + M_{\pi}r) e^{-M_{\pi} r}
\right].
\end{equation}
Above steps result in a model expression for $\sigma_{sea}(t)$:
\begin{eqnarray}
\label{SCFF}
\sigma_{sea}(t) = 
M_{\pi}^2 f_{\pi}^2
\int_{0}^{\infty} d^3{\rvec} j_0(r \sqrt{-t})
\left[ \ds \frac{2 \xi^2}{1+\xi^2}
\right], 
\end{eqnarray}
where $\xi = (R^2_s/r^2) (1+M_{\pi} r) e^{-M_{\pi} r}$.
It follows that due to the partial cancellation of additional contributions
from discrete levels and second order
continuum terms in the $\nabla U$ expansion,
the  accuracy of Eq.~(\ref{SCFF}) in the limiting case
$\sigma_{sea}(t) \to \sigma(t)$ is $\mathcal{O}(15\%)$.
The latter
corresponds to the extreme case that the nucleon scalar quark density is formed
exclusively by the Dirac-sea (pion cloud).
With the expected accuracy Eq.~(\ref{SCFF}) reproduces all features of
the scalar form factor observed in exact $\chi$QSM calculations and
also in $\chi$PT and compares well to lattice QCD
results. Eq.~(\ref{SCFF}) results in pion-nucleon sigma term
$\sigma_{\pi N} = \sigma(0) \simeq 68$~MeV, which is consistent
with empirical information~\cite{Pavan:2001wz,Olsson:1999jt}. 
The analytical simplicity of
Eq.~(\ref{SCFF}) is a big advantage of this method.
We refer to Ref.~\cite{Schweitzer:2003sb} for details and use Eq.~(\ref{SCFF})
for our numerical calculations.
In addition, consider the pion mass expansion of the profile function
$\mathcal{P}(r)$, Eq.~(\ref{ArctPrMpi})
\begin{equation}
\label{ArctPrMpi0}
\mathcal{P}(r,M_{\pi})
= -2 \arctan\left(\frac{R_s^2}{r^2}\right)
+ \mathcal{O}(M_{\pi}^2).
\end{equation}
In the $\chi$QSM the soft pion limit,
$\mathcal{P}(r,M_{\pi} \to 0)$, is known to  approximate well both the
self-consistent profile and Eq.~(\ref{ArctPrMpi}).
Following the method of
Ref.~\cite{Schweitzer:2003sb}, we find that in this limit
$\sigma(t)$ takes an even simpler form 
\begin{equation}
\label{SCFFAN}
\sigma(t) = \frac{4 {\pi}^2}{\sqrt{2}}~  M_{\pi}^2 f_{\pi}^2 R_s^3
~j_0(R_s \sqrt{{-t}/{2}})
~e^{- R_s \sqrt{{-t}/{2}}}.
\end{equation}
If one would expand Eq.~(\ref{SCFF}) in Taylor series around $M_{\pi}=0$,
then Eq.~(\ref{SCFFAN})
corresponds to the leading analytic $\sim M_{\pi}^2$
contribution   to the scalar
form factor or to the $\pi N$  sigma term, $\sigma_{\pi N}$, 
defined by Eq.~(\ref{SCFF}).
The quality of this representation will be discussed later. We only
note that, a simple form of Eq.~(\ref{SCFFAN}) 
is very successful, at least for our
 considerations, because it allows us to express all our final
results for the $r$-space CrTPE potential
 in terms of elementary functions.
 Furthermore, Eq.~(\ref{SCFFAN})
accounts for all details of the CrTPE force observed here and
compares well with direct numerical calculation obtained with
Eq.~(\ref{SCFF}).

\section{Results and discussion}

\subsection{The central CrTPE force: tree-level {\it vs}
unitary $\pi \pi$ scattering amplitude}

\begin{figure}[t]
\includegraphics[clip=true,width=0.72\linewidth]{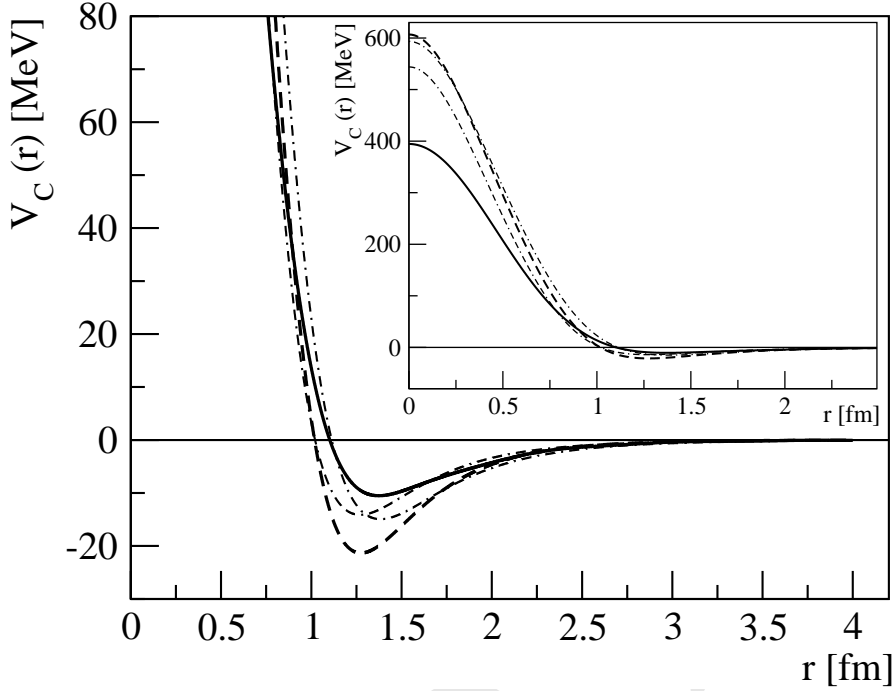}
\caption{\label{CentralTPE} \footnotesize The central CrTPE potential in
                coordinate space obtained with tree level (dashed  curve) and
                unitarized (solid curve) $\pi \pi$ scattering
                amplitude. The dot-dashed and
                dash-dash-dotted curves correspond to the
                soft pion limit with
                $M_q=350$~MeV and
                $M_q=410$~MeV, respectively.
                The insert shows the entire
                structure of the $NN$ potential.}
\end{figure}

In the discussion of our results we start from
 the scalar-isoscalar central $NN$ potential
 in coordinate space, which is given by the Fourier transform
 of the first term in
 Eq.~(\ref{TNNNonRel}). Our results for $\mathcal{V}_{C}(r)$ with
 the scalar form factor $\sigma(t)$ defined by Eq.~(\ref{SCFF}) and
 the tree level $\pi \pi$ correlation function,
 Eq.~(\ref{VPiPi00}), are shown in Fig.~\ref{CentralTPE} (dashed curve).  The
 central potential $\mathcal{V}_{C}(r)$ shows a short-range repulsive
 core with a maximum at the origin, $\mathcal{V}_{C}(r=0) \simeq
 600$~MeV (see insert of Fig.~\ref{CentralTPE}), 
 and is supplemented by a moderate attraction with a
 minimum, $\mathcal{V}_{C} \simeq - 20$~MeV, at intermediate
 distances $r>1$ fm. This behavior is different from the
 conventional picture provided by the  effective $\sigma$ exchange
 in OBE models and also
 from results obtained in the heavy baryon
 $\chi$PT~\cite{Kaiser:1998wa}. In Ref.~\cite{Kaiser:1998wa} the
 effect of the CrTPE in the
 scalar-isoscalar channel is very weak and repulsive.
 The latter means, that the $t$-dependence of the vertex
 functions or functional form of the scalar form factor
 is very important
 for determining the precise strength and behavior of the 
 potential in all ranges
 of distances. At the same time one can find a remarkable agreement
 between our results and the analysis of
 Ref.~\cite{Oset:2000gn}, where the unitary $\pi \pi$ amplitude
 Eq.~(\ref{TPiPiUnitary}) has been coupled to the nucleon cloud at the one
 pion loop. In Ref.~\cite{Oset:2000gn} the divergent loop-integrals
 were regulated by phenomenological
 vertex form factors. Being sensitive to the
 particular choice of the cut-off mass, the
 resulting scalar-isoscalar central $NN$ potential
 shows a strong repulsion at
 internucleon distances less than 1 fm, and a moderate attraction of  
 $-10(-15)$~MeV, at $r > 1$ fm. 

\begin{figure}[t]
\includegraphics[clip=true,width=0.77\linewidth,angle=0]{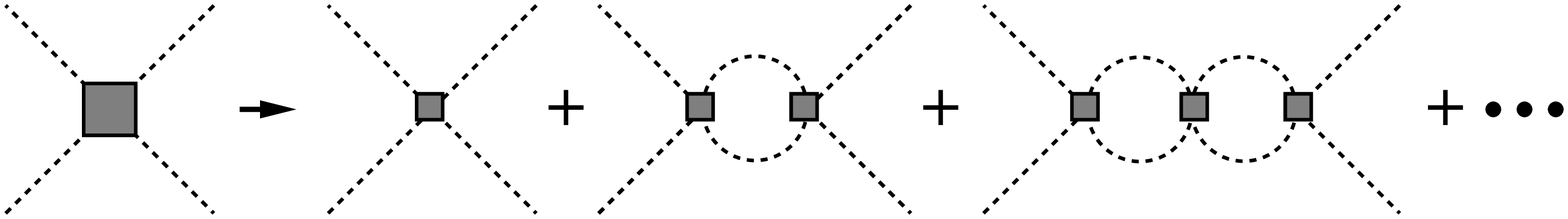}
\caption{\label{UnitPiPiDiagr}
\footnotesize
Unitarization of the 
$\pi \pi$ scattering amplitude.}
\end{figure}

To make our results
 comparable with that from Ref.~\cite{Oset:2000gn} we unitarize 
 the $\pi \pi$ scattering
 amplitude, Fig.~\ref{UnitPiPiDiagr}. 
 It can be done by the following substitution~\cite{Oller:1997ti}
\begin{equation}
\label{TPiPiUnitary}
\tilde{V}^{L=I=0}_{\pi \pi} = -
\left[
{\ds {f^2_{\pi}}\left({ s - \frac{M^2_{\pi}}{2}}\right)^{-1} +
  \ds  G_{\pi \pi}(s)} \right]^{-1},
\end{equation}
where $G_{\pi \pi}(s)$ is a dimensionally regularized two-pion
loop function (see e.g. Ref.~\cite{Oset:2000gn})
\begin{equation}
\label{GREG}
G_{\pi \pi}(s) = \frac{1}{(4 \pi)^2} \left[ -1 + \ln
  \frac{M_{\pi}^2}{\mu^2} + \sigma \ln \frac{\sigma +1}{\sigma -1} \right].
\end{equation}
Here $\mu=1.1$~GeV is the regularization mass fitted to the $\pi
\pi$ phase shifts. The value of $\sigma$ in Eq.~(\ref{GREG}) is given by
$\sigma = \sqrt{1 - \frac{4 M_{\pi}^2}{s}}$.
Eq.~(\ref{GREG}) is analytic in space-like and time-like regions
and  for $s > 4 M_{\pi}^2$ it develops an
imaginary part,
since $\sigma-1<0$ and $\ln \frac{\sigma +1}{\sigma -1} =
\ln \frac{\sigma +1}{1 - \sigma} + i \pi$. 
For $s\to t<0$ the $\log$
behaves smoothly.
Because Eq.~(\ref{TPiPiUnitary}) contains a pole
  in the $s$-channel around $M_{\pi \pi} - i \Gamma/2 \simeq 450 - i 221$~MeV,
  we restore the genuine relation to
  the $\sigma$-meson exchange, which now enters in our formalism as
  dynamical resonance in the $\pi \pi$ system.
  The main effect of the unitary $\pi \pi$ scattering amplitude
  ($\sigma$ meson)
  in the $t$-channel is that it
  makes the repulsion $\simeq 400$~MeV and attraction $\simeq -
  10$~MeV softer (solid curve in Fig.~\ref{CentralTPE}). However it does not
  change the general features of the $NN$ force already obtained at
  tree level. Interestingly, the latter effect is opposite to the 
  role played by the the unitarization 
  in the $s$-channel, where it leads to an
  enhancement of the interaction strength between pions.

Another point is the relation of 
the effective $\sigma$ meson exchange of OBE models to  the
unitary scalar-isoscalar CrTPE force constructed here. 
In Ref.~\cite{Riska:2002vn} the volume integrals were used
to extract the strength of scalar-isoscalar components of
phenomenological $NN$ models and to relate them to the effective
$\sigma$ exchange. It was shown that if the mass of the
$\sigma$ is taken to be $M_{\sigma} = 600$~MeV  then the value of
the ${\sigma NN}$ coupling constant  for
  these interactions would range between $g_{\sigma NN}=7.6$~(Paris),
  9.0~(Argonne V18), 9.8~(Nijmegen) and 11.2 (CD-Bonn).
It follows from Ref.~\cite{Oset:2000gn}, that
 the use of the unitary $\pi \pi$ scattering amplitude Eq.~(\ref{TPiPiUnitary})
 results in an
 effective  ${\sigma NN}$ coupling constant which can be expressed as
\begin{equation}
\label{SigmaNN}
g_{\sigma NN} \simeq \sqrt{6} \, V(t=0) \left(M_{\sigma}^2 -
{M_{\pi}^2}/{2} \right) /f_{\pi}.
\end{equation}
Here, $V(t)$ is a vertex function which accounts for the coupling of the
$\pi \pi$ correlation function to the nucleon pion cloud, and $M_{\sigma}$
is an empirical $\sigma$ meson mass. 
Note that the use of $M_{\sigma} =
600$~MeV in this case
is not entirely correct, because this value does not represent the
actual pole of Eq.~(\ref{TPiPiUnitary}). With the quoted
value of $V(t)$ at $t=0$ \cite{Oset:2000gn}: $V(0)\simeq 0.1 \times
10^{-2}$~MeV$^{-1}$ and the $\sigma$ mass (pole) $M_{\sigma} \simeq 450$~MeV
the resulting coupling is $g_{\sigma NN} \simeq 5$, which has the
right order of magnitude. In our case,
  note that the vertex function $V(t)$ as
provided by Ref.~\cite{Oset:2000gn} is equivalent to our
$\Gamma_N^{+}(t)=V(t)$ in Eq.~(\ref{VNNGamma}). By this, we
rewrite Eq.~(\ref{SigmaNN}) and relate the $\sigma NN$ coupling
constant to the $\pi N$ sigma term $\sigma_{\pi N}$
\begin{equation}
\label{GSigmaSiNN}
g_{\sigma NN} \simeq  \sqrt{\frac{2}{3}}\,  \sigma_{\pi N} \left(M_{\sigma}^2 -
M_{\pi}^2/2\right) /(f_{\pi} M_{\pi}^2).
\end{equation}
Eq.~(\ref{SCFF}) results in $\sigma_{\pi N} \simeq 68$~MeV and our
value for the $V(t=0)$ is given by: $V\simeq 0.12 \times
10^{-2}$~MeV$^{-1}$. With $M_{\sigma} = 450$~MeV our result is
$g_{\sigma NN} \simeq 6.1$. An artificial increase of the $\sigma$ mass to
$M_{\sigma}=600$~MeV, used in analysis of Ref.~\cite{Riska:2002vn},
results in $g_{\sigma NN} \simeq 11$ which is just in the
range of values from phenomenological models~\cite{Riska:2002vn}.

Using the equivalence between
$\Gamma_N^{+}(t)$ and $V(t)$ we independently verify  the
theoretical consistency  of Ref.~\cite{Oset:2000gn}. Indeed,
with $V(0)$ 
obtained in Ref.~\cite{Oset:2000gn} the resulting $\pi N$ sigma
term is $\sigma_{\pi N} \simeq 57$~MeV and is
within the range of empirical
values~\cite{Pavan:2001wz,Olsson:1999jt}. 
It means that values of cut-off masses used in 
Ref.~\cite{Oset:2000gn} are reasonable and consistent with our
results and with empirical information.
There is also consistency between the large-$N_c$ behavior of
Eq.~(\ref{GSigmaSiNN}) and large-$N_c$ QCD expectations, where a
scalar-meson $NN$ coupling constant is
$\sim \mathcal{O}(\sqrt{N_c})$~\cite{Kaplan:1996rk}.
Because the $1/N_c$ expansion of
the $\pi N$ sigma term follows the large-$N_c$ behavior of
the scalar form factor $\sigma_{\pi N} \sim \mathcal{O}(N_c)$, the
pion decay constant $f_{\pi} \sim \mathcal{O}(1/\sqrt{N_c})$,
$M_{\pi} \sim \mathcal{O}(1)$  and 
$M_{\sigma} \sim \mathcal{O}(1)$~\cite{Pelaez:2003dy}
we get
\begin{equation}
g_{\sigma NN} \sim \mathcal{O}(\sqrt{N_c}),
\end{equation}
in agreement with large-$N_c$  counting rules~\cite{Kaplan:1996rk}.

The noncommutativity of large-$N_c$ and chiral 
limits is a well-known phenomenon~\cite{Dashen:1993jt}. 
The important physics behind this is
the role of the $\Delta(1232)$ isobar. In the large-$N_c$ limit 
nucleon and $\Delta$ are degenerate  with
mass difference $M_{\Delta} - M_{N}  \sim \mathcal{O}(1/N_c)$.
As a consequence the contribution of $\Delta$ states
is implicitly included in solitonic configurations and 
their effect on scalar-isoscalar quantities is twice that of nucleon
states~\cite{Cohen:1992uy,Schweitzer:2003sb}. 
Note that, therefore the leading non-analytic 
contribution to $\sigma_{\pi N}$, which is  
$27 g_{A}^2 M_{\pi}^{3}/(64 \pi f_{\pi}^2)$ obtained by 
expanding Eq.~(\ref{SCFF}), 
is exactly three times larger than the corresponding value in 
$\chi$PT including nucleons and pions only. 
Being common for generic classes of hedgehog
models~\cite{Cohen:1992uy}, 
this result can be explained by the presence of the intermediate
$\Delta$ states in chiral loops. 
If one goes beyond the strict 
large-$N_c$ limit the effect of 
finite $N-\Delta$ mass splitting must be taken into account,
the magnitudes of the resulting corrections are known~\cite{Cohen:1992uy}.

\subsection{Saturation of the LEC $c_1$}
Here we use constraints imposed by chiral symmetry to
check the consistency of the present approach. Consider the
LEC $c_1$ from $\chi$PT which is related to the $\pi N$ sigma term 
$\sigma_{\pi N}$~\cite{Bernard:1995dp}. 
The phenomenological interpretation of its value is based on
the strongly coupled scalar-isoscalar meson exchange which can
completely saturate $c_1$, if~\cite{Bernard:1996gq} 
\begin{equation}
\frac{M_{\sigma}}{\sqrt{g_{\sigma NN}}} =180~ \mbox{MeV}.
\end{equation} 
It is interesting to note that the effective $\sigma$ meson in the Bonn
potential~\cite{BonnPT} with $M_{\sigma}=550$~MeV and
$g_{\sigma NN}^2/(4 \pi) = 7.1$ 
respects this condition   
resulting in ${M_{\sigma}}/{\sqrt{g_{\sigma NN}}} 
\simeq 179$~MeV.
In our case, using the pole  of 
Eq.~(\ref{TPiPiUnitary}), i.e. $M_{\sigma} = 450$~MeV, and
$g_{\sigma NN} = 6.1$, Eq.~(\ref{GSigmaSiNN}), the saturation
condition is 
\begin{equation}
\frac{M_{\sigma}}{\sqrt{g_{\sigma NN}}} \simeq 182~ \mbox{MeV}.
\end{equation} 
which is in agreement with the value demanded from the scalar 
meson resonance saturation of the LEC $c_1$.

\begin{figure}[t]
\includegraphics[clip=true,width=0.77\linewidth]{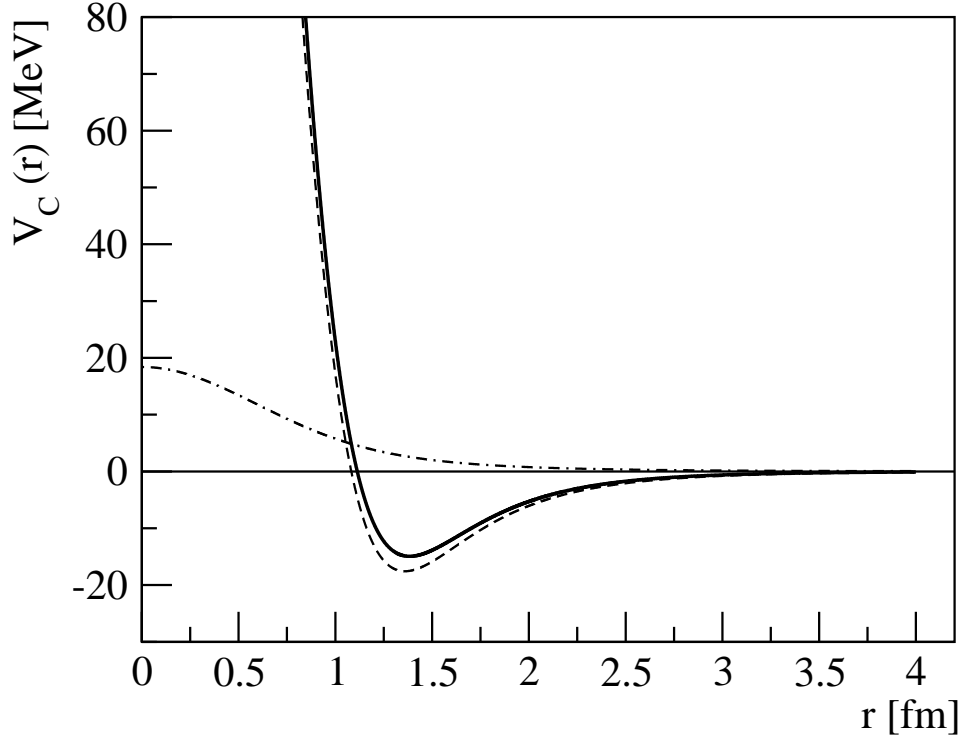}
\caption{\label{CentralAn} \footnotesize
               The central CrTPE potential
               in the soft pion limit (solid curve).
               The parts non-vanishing and vanishing
               in the chiral limit are shown by
               the dashed and dot-dashed curves, respectively.}
\end{figure}

\subsection{Analytical formulae for the central force}
The consideration of the central part can be continued 
 in the soft pion limit where $\sigma(t)$ is given by Eq.~(\ref{SCFFAN}).
 In this case the central CrTPE force in the coordinate space representation
 can be expressed in terms of elementary functions. First, 
 decompose the tree level $\pi \pi$ correlation function defined by 
 Eq.~(\ref{VPiPi00}) into two terms,
 $\tilde{V}^{L=I=0}_{\pi \pi} = \tilde{V}^{(1)}_{\pi \pi}
 + \tilde{V}^{(2)}_{\pi \pi}$:
\begin{equation}
\label{VPiPiDecomp}
\tilde{V}^{(1)}_{\pi \pi} = -t /f_{\pi}^2, ~~~
\tilde{V}^{(2)}_{\pi \pi} = M_{\pi}^2/2 f_{\pi}^2.
\end{equation}
By this, the CrTPE force reads as
\begin{equation}
\label{VNN12} \mathcal{V}_{NN}(t) = \mathcal{V}^{(1)}_{NN}(t) +
\mathcal{V}^{(2)}_{NN}(t).
\end{equation}
In the soft pion limit, due to the factor $1/M_{\pi}^2$ entering
 Eq.~(\ref{TNNNonRel}), the first term $\mathcal{V}^{(1)}_{NN}$
 does not contain any pion mass dependence and does not vanish in
 the chiral limit (non-vanishing part). The second term
 $\mathcal{V}^{(2)}_{NN}$ which follows $\tilde{V}^{(2)}_{\pi \pi}$
 and contains an additional $M_{\pi}^2$ generates the symmetry
 breaking part of the $NN$ force. With Eqs.~(\ref{SCFFAN})
 and~(\ref{VPiPiDecomp})  the Fourier transform of
 Eq.~(\ref{VNN12}) can be carried out explicitly and after lengthy
 calculations (here we list our final results only) the central
 CrTPE force is given in coordinate space  by following
 expressions
\begin{eqnarray}
\label{VC1} \mathcal{V}^{(1)}_{C}(r) &=&  \left( \frac{64\,\sqrt{2}\,
\pi^2}{3} \right) f_{\pi}^2 R_s
\hspace{4.cm} \\
&&\times \left[ {\frac {80+8\,\tilde{\xi}+3\,\tilde{\xi}^{2}
-3\,\tilde{\xi}^{3}}{\left (2+\tilde{\xi}\right )^{2}\left
(4+\tilde{\xi} +2\,\sqrt{2 \tilde{\xi}} \right )^ {2}\left
(4+\tilde{\xi}-2\, \sqrt{2 \tilde{\xi}} \right )^{2}}} \right] \nonumber
\end{eqnarray}
\begin{eqnarray}
\label{VC2a} \mathcal{V}^{(2)}_{C}(r,\tilde{\xi}\leq4) &=&
\frac{M_{\pi}^2}{r}
\left( \frac{4\, \pi^2}{3}  \right) f_{\pi}^2 R_s^4
\hspace{2.5cm}
 \\ &&\times
\left[ \arctan\left(\sqrt{\frac{\tilde{\xi}}{2}}\right) -
\frac{1}{2} \arctan\left(\frac{ 2 \sqrt{2\tilde{\xi} }}{4 -
    \tilde{\xi}}
\right) \right] \nonumber \\
\label{VC2b} \mathcal{V}^{(2)}_{C}(r,\tilde{\xi}>4) &=&
\frac{M_{\pi}^2}{r}
\left( \frac{4\, \pi^2}{3}  \right) f_{\pi}^2 R_s^4
\hspace{2.5cm}
 \\ &&\times
\left[ \arctan\left(\sqrt{\frac{\tilde{\xi}}{2}}\right) -
\frac{1}{2} \left\{ \pi - \arctan\left(\frac{ 2 \sqrt{2\tilde{\xi}
}}{\tilde{\xi} - 4} \right) \right\} \right] \nonumber
\end{eqnarray}
where ${\tilde{\xi}}= r^2/R^2_s$ and the superscripts (1) and (2)
refer to
the non-vanishing part and the symmetry breaking part, respectively. Note
that  Eqs.~(\ref{VC2a}) and ~(\ref{VC2b}) are defined for
$\tilde{\xi} \leq 4$ and $\tilde{\xi} > 4$, respectively.

At small separation scales Eqs.~(\ref{VC1}) and~(\ref{VC2a}) can be
expanded in Taylor series  around $r\to 0$, and  up to
$\mathcal{O}( {r}^{4})$ the leading terms are given by:
\begin{eqnarray}
\label{VC1To0} \mathcal{V}^{(1)}_{C}(r)\Big|_{r \to 0} &=&
\frac{5\sqrt{2} \pi^2}{3}~  f_{\pi}^2 R_s \, - \frac{3
\pi^2}{\sqrt{2}}~  \frac{f_{\pi}^2}{R_s}\, r^2 +\mathcal{O}(
{r}^{4}), \nonumber \\ \\
\label{VC2To0} \mathcal{V}^{(2)}_{C}(r)\Big|_{r \to 0} &=&
\frac{\sqrt{2} \pi^2}{3}  f_{\pi}^2 M_{\pi}^2   R_s^3 \nonumber \\
&&- \frac{5 \sqrt{2} \pi^2}{36}  \ f_{\pi}^2 M_{\pi}^2 R_s r^2 +
\mathcal{O}(r^4).
\end{eqnarray}
The strength of the central force is defined
by the sum of Eqs.~(\ref{VC1To0}) and~(\ref{VC2To0})  
at $r=0$: $\mathcal{V}_C(0) =
\mathcal{V}^{(1)}_{C}(0) + \mathcal{V}^{(2)}_{C}(0)$
\begin{eqnarray}
\label{VCStrength} \mathcal{V}_{C}(r)\Big|_{r=0} &=& \frac{5
\sqrt{2} \pi^2}{3} \left(1
  +  \frac{1}{5} M_{\pi}^2 R_s^2
\right)  f_{\pi}^2 R_s.
\end{eqnarray}
For asymptotically large $r\to \infty$ (up to
$\mathcal{O}(r^{-8})$) the interaction is dominated by
molecular-like $NN$ forces
\begin{eqnarray}
\mathcal{V}^{(1)}_{C}(r) \Big|_{r \to \infty} &=&
-64 \sqrt {2} \, \pi^2 f_{\pi}^2 R_s^7 {\frac
  {1}{{r}^{6}}}+\mathcal{O}({r}^{-8}),
\\ \nonumber \\
\mathcal{V}^{(2)}_{C}(r)\Big|_{r \to \infty} &=& \frac{8
\sqrt{2} \pi^2}{3}   M_{\pi}^2  f_{\pi}^2 R_s^7 \frac{1}{r^4}  \\
&&- \frac{16 \sqrt{2} \pi^2}{3}   M_{\pi}^2  f_{\pi}^2 R_s^9
\frac{1}{r^6} + \mathcal{O}(r^{-8}). \nonumber
\end{eqnarray}

As seen from above expressions, $\mathcal{V}^{(1)}_{C}$ and
$\mathcal{V}^{(2)}_{C}$ do not vanish at small separations and
both are repulsive. At asymptotically large $r$
$\mathcal{V}^{(1)}_{C}$ becomes attractive, signaling that it
should cross the zero. Contrary, the symmetry breaking part,
$\mathcal{V}^{(2)}_{C}$, stays repulsive for all distances and as
a result of $\mathcal{V}^{(2)}_{C}/\mathcal{V}^{(1)}_{C} \sim
M_{\pi}^2 R_s^2/5 \ll 1$ it is 
suppressed relative to $\mathcal{V}^{(1)}_{C}$. The above
qualitative arguments are supported by full analytical
representations given by Eqs.~(\ref{VC1}-\ref{VC2b}). The resulting
central $NN$ potential is
shown in Fig.~\ref{CentralAn}. It is clear from
Fig.~\ref{CentralAn} that the  part $\mathcal{V}_C^{(1)}(r)$, which is
non-vanishing in the chiral limit (dashed curve), is dominant and
fully drives the generic behavior of the central CrTPE force
(solid line). The chiral part $\mathcal{V}_C^{(2)}(r)$,
which is vanishing in the chiral limit (dot-dashed curve),
is repulsive for all distances with a strength of about $\simeq
20$~MeV at $r=0$. Comparison with numerical results from the
previous section reveals that at intermediate distances (dip
region) the total CrTPE force obtained in the soft pion limit
 (shown by the dot-dashed curve in Fig.~\ref{CentralTPE}) lies
in-between the tree level and unitarized curves. At small
separation scales (see insert of Fig.~\ref{CentralTPE}) the
integral Eq.~(\ref{SCFF}) and analytical Eq.~(\ref{SCFFAN})
representations for the scalar form factor work equally well, and
the analytical and numerical tree level results are very close to
each other.

Up to now, we used the common value of the dynamical
quark mass $M_q=350$~MeV motivated by the instanton
phenomenology. We have to mention that with this value the
representation of $\sigma(t)$
Eq.~(\ref{SCFFAN}) overestimates the value of the $\pi N$
sigma term obtained with Eq.~(\ref{SCFF}) and the latter can be
reproduced by a small variation of $M_q$. Note that,
in the $\chi$QSM
the dynamical quark mass is usually treated as a free parameter varying
in the range $350 < M_q <450$~MeV. So it is important to
check whether our results are sensitive to the variation of $M_q$.
Certainly, due to Eq.~(\ref{VCStrength}) and the parametric
smallness of the term $M_{\pi}^2 R_s^2/5 \ll 1$ the strength of the
repulsive core is proportional to $\sim 1/M_q$. Interestingly, the
strength of the intermediate range attraction actually is
insensitive to variations of quark mass. 
In Fig.~\ref{CentralTPE} we show the
results for $M_q \simeq 410$~MeV (dash-dash-dotted
curve) which reproduces the same value of the sigma term $\sigma_{\pi N}$
as Eq.~(\ref{SCFF}) with $M_q=350$~MeV. The dip region is
unaffected by these variations with a small coherent shift towards
shorter distances.  The same weak quark mass dependence is found in
numerical calculations. Summarizing, the soft pion limit
provides an accurate representation of the central CrTPE force,
compares well with direct numerical calculations and accounts for
all details of the tree level and unitary forces obtained in the
previous section.

\subsection{The spin-orbit CrTPE force}
\begin{figure}[t]
\includegraphics[clip=true,width=0.77\linewidth]{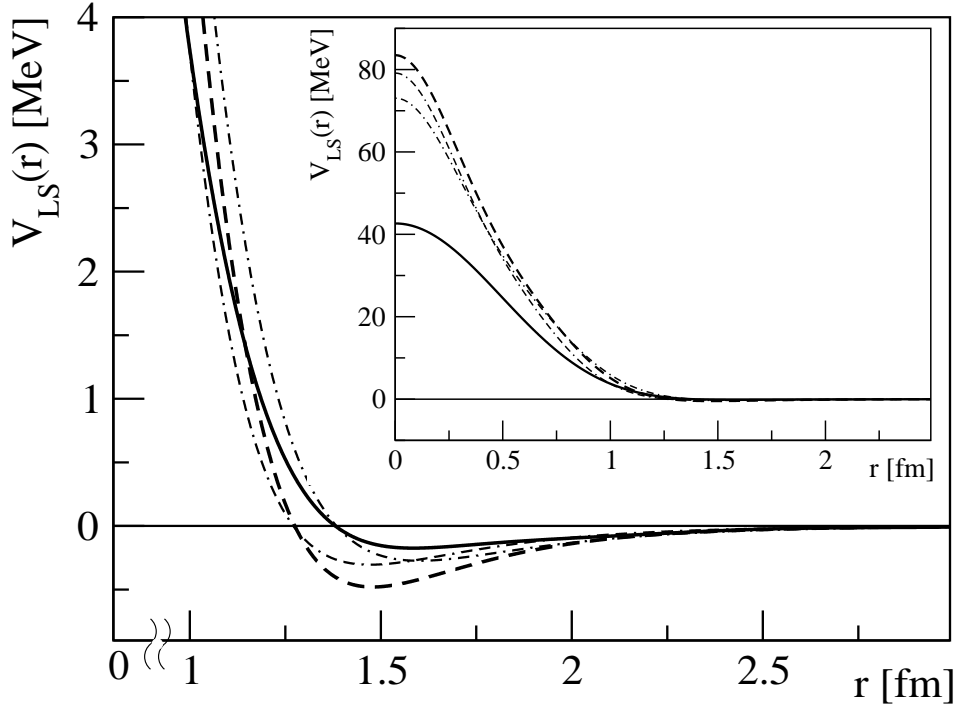}
\caption{\label{SpinOrbitCrTPE} \footnotesize
                The spin-orbit part of the CrTPE potential
                in coordinate space. Notations for the
                curves are the same as in Fig.~\ref{CentralTPE}.
                The insert shows the entire
                structure of the $LS$ potential.} 
\end{figure}

In coordinate space the spin-orbit part of the scalar-isoscalar CrTPE force is
related to the central potential by Eq.~(\ref{VLSCOORD}). According
to the large-$N_c$ analysis given in section IV,
the $LS$ interaction should be much smaller than the
central part because it is suppressed relative to $\mathcal{V}_C(t)$
by $\sim 1/N_c^2$.
Our numerical results for the $LS$ potential are shown in
Fig.~\ref{SpinOrbitCrTPE}. Indeed, the $r$-space
behavior of $\mathcal{V}_{LS}(r)$ with the tree level
$\pi \pi$ scattering amplitude (dashed curve in
Fig.~\ref{SpinOrbitCrTPE}) is formally similar to the central
potential (see the insert of Fig.~\ref{CentralTPE}). But contrary
to the $\mathcal{V}_C(r)$, we find a sizable but much weaker repulsive core
$\simeq 80$~MeV (insert of Fig.~\ref{SpinOrbitCrTPE})
and a very small attraction $\simeq 0.5$~MeV at
intermediate distances $r>1.2$~fm.  The
effect of the unitarization of the $\pi \pi$ scattering amplitude
(solid curve in Fig.~\ref{SpinOrbitCrTPE}) plays here the same
role as for the central potential and makes the $LS$ force softer
reducing the strength of the potential by the factor two.
Note, that in OBE models, the
$LS$ interaction induced by effective $\sigma$ exchange generates  an
attractive spin-orbit force. In our case the effect is opposite
and leads to a repulsive $LS$ potential in coordinate space.
In Ref.~\cite{Oset:2000gn} the spin-orbit potential was not
calculated (though it could be easily done) and hence direct
comparison is not possible. We only note, that the results should
be similar because of Eq.~(\ref{VLSCOORD}) and because of the similar
behavior of $\mathcal{V}_C(r)$ in both approaches.

At this level the comparison with other models is also useful. For
example, in the standard Skyrme model~\cite{Skyrme:vh}, which
includes the non-linear $\sigma$ model Eq.~(\ref{Lpipi}) and a
stabilizing fourth-order term, the ``wrong sing'' of the isospin
independent spin-orbit potential is a long standing
problem~\cite{Riska:nk,Abada:1996ux}. Several attempts have
already been done in this framework to generate the needed attractive
spin-orbit force, extending the
standard Skyrme Lagrangian by including sixth-order derivative
terms. But the final results differ, varying from
repulsive~\cite{Abada:1996ux}
to attractive~\cite{Riska:1989fw} spin-orbit potential. Our results support the
existence of a repulsive $LS$ interaction in the
isoscalar channel.

\begin{figure}[t]
\includegraphics[clip=true,width=0.77\linewidth]{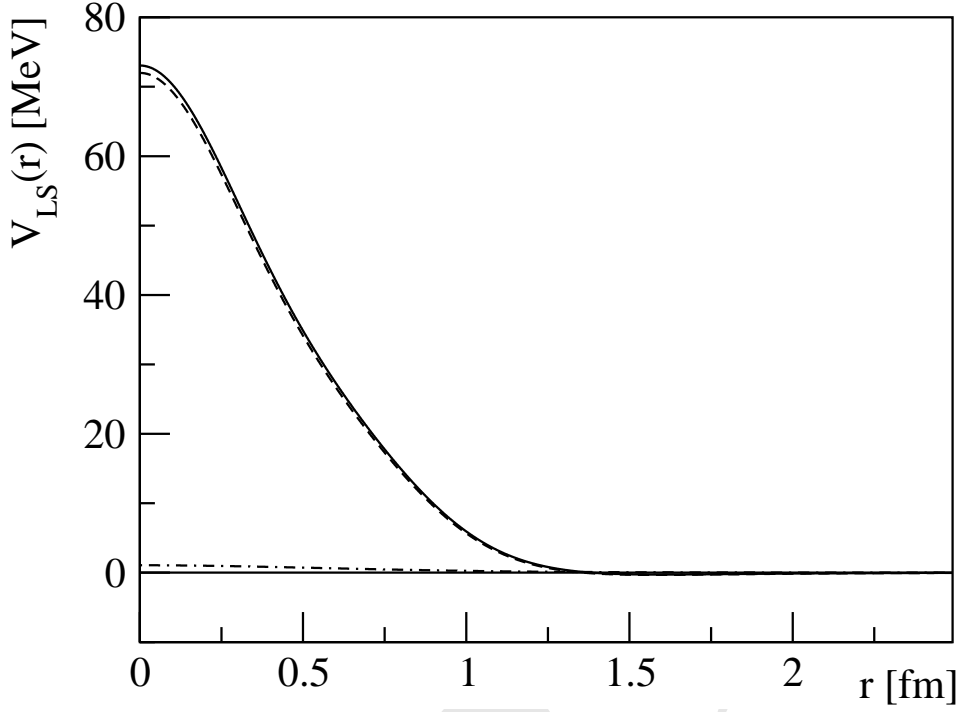}
\caption{\label{SpinOrbitCrTPEAn} \footnotesize
                The spin-orbit part of the CrTPE potential
                in the soft pion limit. Notations for
                the curves are the same as in Fig.~\ref{CentralAn}.}
\end{figure}

\subsection{Analytical formulae for the spin-orbit force}
Here we show an analytical representation
of the spin-orbit potential in coordinate space in terms of elementary functions.
Using Eqs.~(\ref{VLSCOORD}) and~(\ref{VC1}), the $LS$ part
$\mathcal{V}^{(1)}_{LS}$, which is non-vanishing in the chiral
limit, reads
\begin{eqnarray}
\label{VLS1} \mathcal{V}^{(1)}_{LS}(r) &=&
\left(\frac{64\,\sqrt{2}\, \pi^2}{3}\right) \frac{f_{\pi}^2}{R_s M_N^2}
\hspace{3.5cm} \\
&&\times \left[ {\frac
{2304+576\,\tilde{\xi}+816\,{\tilde{\xi}}^{2}+100\,{\tilde{\xi}}^{3
}-6\,{\tilde{\xi}}^{4}-9\,{\tilde{\xi}}^{5}}{\left
(2+\tilde{\xi}\right )^{3}\left ( 4+{\tilde{\xi}}+2\,\sqrt {2
\tilde{\xi}}\right )^{3} \left (4+{\tilde{\xi}}-2\,\sqrt {2
\tilde{\xi} } \right )^{3}}} \right] \nonumber
\end{eqnarray}
Its symmetry breaking  counterpart $\mathcal{V}^{(2)}_{LS}$ 
vanishes in the chiral limit,
 $M_{\pi} \to 0$, and  is obtained simply from
Eqs.~(\ref{VLSCOORD}),~(\ref{VC2a}) and~(\ref{VC2b})
\begin{eqnarray}
\label{VLS2} \mathcal{V}^{(2)}_{LS}(r) &=& \frac{1}{2 M_N^2}
\frac{1}{r^2} \mathcal{V}_{C}^{(2)}(r)  \hspace{3.9cm} \\ && -
\frac{M_{\pi}^2}{r^2} \left( \frac{4\, \sqrt {2} \,\pi^2}{3} \right)
\frac{f_{\pi}^2 R_s^3}{M_N^2}
\left[ \frac{4\,-3\,\tilde{\xi}}{\left
(16\,+\tilde{\xi}^{2}\right )\left (2\,+\tilde{\xi}\right )} \right]
\nonumber
\end{eqnarray}
where for $\tilde{\xi}=r^2/R_S^2 \leq 4$ and $\tilde{\xi} > 4$ the
$\mathcal{V}^{(2)}_C(r)$ is given by Eqs.~(\ref{VC2a}) and~(\ref{VC2b}),
respectively. The Taylor expansion around $r \to 0$ up to
$\mathcal{O}(r^4)$ reads
\begin{eqnarray}
\label{VLS1RTO0} \mathcal{V}^{(1)}_{LS}(r)\Big|_{r\to 0} &=&
\frac{3\,{\pi }^{2}}{\sqrt{2}}\,{\frac
{f_{\pi}^{2}}{R_s\,{M}_N^{2}} } \\ &&-{\frac {15\,\sqrt {2}{\pi
}^{2}}{8}}\,{\frac {
    f_{\pi}^{2}}{R^3_s\,M_N^{2}}}{r}^{2}+\mathcal{O}\left
({r}^{4}\right), \nonumber \\
\label{VLS2To0} \mathcal{V}^{(2)}_{LS}(r)\Big|_{r \to 0} &=&
\frac{5
\sqrt{2}\,{\pi}^2}{36}  f_{\pi}^2 R_s \frac{M_{\pi}^2 }{M_N^2} \\
&&- \frac{3 \sqrt{2}\,{\pi}^2}{40}  f_{\pi}^2 \frac{1}{R_s}
\frac{M_{\pi}^2 }{M_N^2} r^2 + \mathcal{O}(r^4),
 \nonumber
\end{eqnarray}
and defines the short-range behavior of the $LS$ force with the
total strength at origin
\begin{eqnarray}
\label{VLSStrength} \mathcal{V}_{LS}(r)\Big|_{r \to 0} = \frac{5
\sqrt{2}\,{\pi}^2}{3} \left(\frac{9}{10}+\frac{1}{12} M_{\pi}^2
R_s^2 \right)  \frac { f_{\pi}^2}{R_s\,{M}_N^{2}}.
\end{eqnarray}

In the asymptotic region, $r \to \infty$, the soft two-pion tail
of the $LS$ potential behaves as follows
\begin{eqnarray}
\mathcal{V}^{(1)}_{LS}(r)\Big|_{r\to \infty} &=& -192\,{\frac
{\sqrt {2}{\pi }^{2}}{{M}_N^{2}}} f_{\pi}^{2} R_s^7\,
\frac{1}{r^8}
+ \mathcal{O}({r}^{-10})
\end{eqnarray}
\begin{eqnarray}
\label{VLS2toInfty} \mathcal{V}^{(2)}_{LS}(r)\Big|_{r \to \infty}
&=& \frac{16 \sqrt{2}\,{\pi}^2}{3}  f_{\pi}^2 R_s^7
\frac{M_{\pi}^2 }{M_N^2} \frac{1}{r^6} + \mathcal{O}(r^{-8})
\end{eqnarray}
 The result of Eqs.~(\ref{VLS1})
  and  ~(\ref{VLS2}) with the quark mass $M_q=350$~MeV
  are shown in Fig.~\ref{SpinOrbitCrTPEAn}.
  The soft pion limit well reproduces the generic features of
  the $LS$ force (solid line) and compares well with our numerical
  results  (dot-dashed curve in Fig.~\ref{SpinOrbitCrTPE}).
  Contrary to the central force, even smaller contribution of the chiral
  part $\mathcal{V}^{(2)}_{LS}(r)$ (dot-dashed curve in
  Fig.~\ref{SpinOrbitCrTPEAn}) with repulsive strength at $r=0$ of 
  about $\simeq 0.11$~MeV is
  noted. The  term $\mathcal{V}^{(1)}_{LS}(r)$
  non-vanishing in the chiral limit
  accurately  represents
  the behavior of the $LS$ potential in all ranges of distances and 
  is actually insensitive to the
  variation of quark masses in the intermediate region. The increase
  of the quark mass to $M_q=410$~MeV leads to a small shift
  of the attractive dip towards shorter distances (dash-dash-dotted
  curve in Fig.~\ref{SpinOrbitCrTPE}).
  At short separation scales, Eq.~(\ref{VLSStrength}), 
  $M_{\pi}^2 R_s^2/12 \ll 1$
  and the effect of quark mass is linear $\sim M_q$ and
  opposite to the central force where the strength is driven by $\sim 1/M_q$.

\begin{figure}[t]
\includegraphics[clip=true,width=0.77\linewidth]{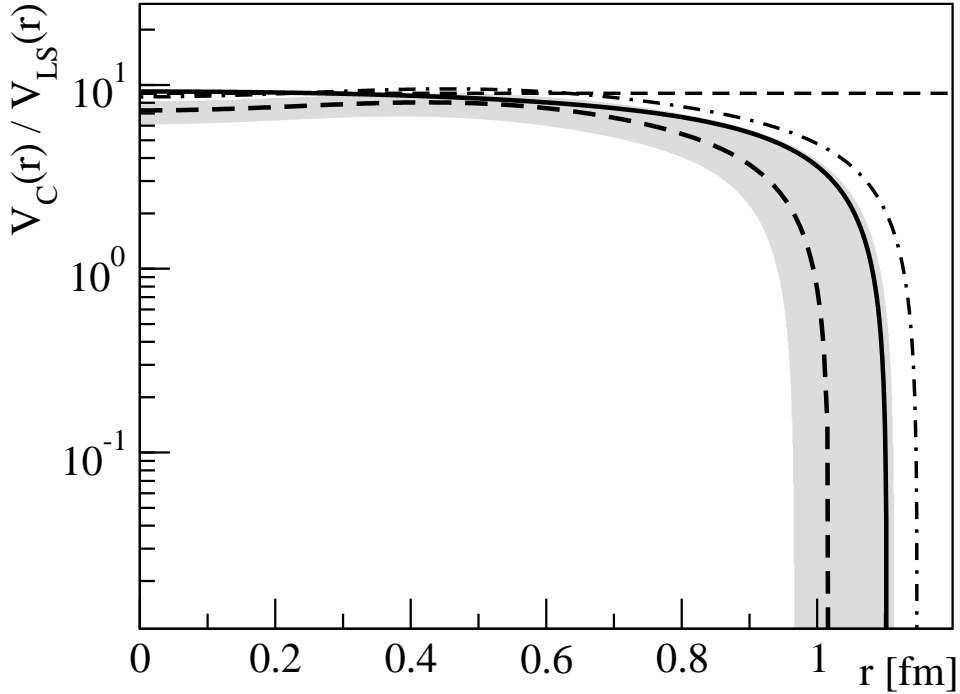}
\caption{\label{RatioVCToVSO} \footnotesize 
               The scaling  of the central CrTPE force
               $\mathcal{V}_{C}(r)$ relative
               to the spin-orbit potential  $\mathcal{V}_{LS}(r)$. The
               numerical results are shown with the tree level  (dashed
               curve) and unitary (solid curve) $\pi \pi$ correlation
               function. The gray area is the soft pion
               limit with quark masses $M_q=350$ (upper limit)
               to ~$410$~MeV (lower limit). The dot-dashed curve
               corresponds the soft pion limit but with $R_s$ related 
               to the $g_A$. The dashed horizontal
               line represents the large-$N_c$ QCD 
               expectation, $1/\mathcal{R}=9.$}
\end{figure}

\subsection{The ratio $V_{LS}/V_{C}$ and large-$N_c$ QCD scaling}
According to large-$N_c$ counting rules discussed in section IV
and Eq.~(\ref{VLStoVSscaling}), one can expect that with the actual number of
colors $N_c=3$ the ratio $V_{LS}/V_{C}$ should scale according to the
following simple relation
\begin{equation}
\label{VLStoVC} 
\mathcal{V}_{LS}/\mathcal{V}_{C} \sim
\mathcal{O}(1/N^2_c) = \mathcal{O}(1/9) \simeq 0.11
\end{equation}
Our results support these expectations and
Eq.~(\ref{VLStoVC}) for the relative strength of scalar-isoscalar
$NN$ components, $\mathcal{R} = \mathcal{V}_{LS}(0)/\mathcal{V}_{C}(0)$,
 is in a good agreement with numerical calculations.
With $\sigma(t)$ defined by Eq.~(\ref{SCFF})
and with the tree level $\pi \pi$ correlation function the ratio is: $
\mathcal{R} \simeq 1/7.26 \simeq 1.14$.
The unitarization of the $\pi \pi$ scattering
amplitude, Eq.~(\ref{TPiPiUnitary}), plays here an important role
and improves this value to
$\mathcal{R} \simeq 1/9.2 \simeq 0.11$. 
Considering the latter value as our parameter free prediction,
it is remarkable that such detailed information about the relative
strength of potential components can be deduced directly from QCD.
Note that the $1/N_c$ expansion is equivalent to the
short-distance expansion~\cite{Witten:1979kh,Beane:2002ab}. 
This is the reason, why Eq.~(\ref{VLStoVC}) 
so accurately represents the short-distance part of the CrTPE force.
At the same time, it is
very difficult to argue that the scaling of scalar-isoscalar
nuclear forces, first predicted by Kaplan and
Manohar~\cite{Kaplan:1996rk}, and the observation which we made
here, are not accidental, because it is very difficult to justify
any phenomenological models at such separation scales where the
baryon number is not well
defined~\cite{PhenSigmaModels,Kaskulov:2002mc} 
and where other
hard QCD processes~\cite{Kaskulov:2003wh}
should play some role. We also recall that the $\chi$QSM itself as
provided by
the action Eq.~(\ref{SMPartFunc}) is valid for the values of the quark
momenta up to the
UV cut-off $\Lambda = 1/\langle \rho \rangle \simeq 600$~MeV or
$r>r_{\Lambda}\simeq 0.3$~fm.
In addition, we have to take into account the limited accuracy of 
the method used here to calculate $\sigma(t)$.
Considering
our findings as an empirical fact, we note that, other models, e.g.
Ref.~\cite{Oset:2000gn,Inoue:2003bk}, should be employed here to
check our statements.
To further support the scaling relation
Eq.~(\ref{VLStoVC}) we show in Fig.~\ref{RatioVCToVSO} the
$r$-dependence of the inverse ratio
$1/\mathcal{R}(r)=\mathcal{V}_{C}(r)/\mathcal{V}_{LS}(r)$.  Indeed, the ratio
$\mathcal{V}_{C}(r)/\mathcal{V}_{LS}(r)$ scales  according
to Eq.~(\ref{VLStoVC}) forming a plateau up to distances $r \simeq
0.8$~fm and fastly decreases beyond.
The plateau covers the region where the
action Eq.~(\ref{SMPartFunc}) is already reliable
and the present approach is well justified.

A good agreement with the scaling relation 
 is also found for the soft pion limit. 
 For $M_q=350$~MeV and $410$~MeV,  
 Eqs.~(\ref{VCStrength}) and (\ref{VLSStrength})
 result in $\mathcal{R} \simeq 0.12$ and $\simeq 0.16$, respectively.
 The radial dependencies of  
 the inverse ratios are shown in
 Fig.~\ref{RatioVCToVSO} by the band, where the upper limit
 corresponds to $M_q=350$~MeV and lower one to $M_q=410$~MeV.
 Additionally, due to the smallness of $M_{\pi}^2 R_s^2$ in
 Eqs.~(\ref{VCStrength}) and (\ref{VLSStrength}), the following 
 relation holds: $\mathcal{R} \simeq (9/10) (M_q^2/M_N^2)$.
 Using Eq.~(\ref{GACoupl}) with $R_{s}=1/M_q$ it can
 be written alternatively
\begin{equation}
\label{RAxial}
1/\mathcal{R} \simeq \frac{5}{6}
\left(\frac{g_{A}}{4 \pi}\right)
\frac{M_N^2}{f_{\pi}^2}.
\end{equation}
 With the empirical value for the axial coupling constant, $g_A \simeq
 1.25$, Eq.~(\ref{RAxial}) results in $\mathcal{R}
 \simeq 0.12$ in good agreement with Eq.~(\ref{VLStoVC}). The
 $r$-dependence of $1/\mathcal{R}$ with the soliton
 size $R_s$ related to the axial coupling constant $g_A$, Eq.~(\ref{GACoupl}), 
is shown in Fig.~\ref{RatioVCToVSO} by the dot-dashed curve.

\section{Summary}
In summary, we have considered the effect of correlated two-pion
 exchange modes on the central and spin-orbit parts of the
 scalar-isoscalar $NN$  interaction. We have coupled the
 $\pi \pi$ correlation function to the scalar form factor of the
 nucleon 
 which we calculated in
 the large-$N_c$ limit in the framework of $\chi$QSM.
 The CrTPE force obtained here confirms an unconventional behavior
 of scalar-isoscalar $\pi \pi$ correlations in the $NN$ interaction
 recently found in unitarized $\chi$PT~\cite{Oset:2000gn}.
 For the central $NN$ potential with, both, the tree level and
 unitary $\pi \pi$ correlation functions we find a strong repulsive
 core at short ranges and a moderate attraction at intermediate
 distances. This result indicates that strong repulsive interactions can be
 generated by the pion-pion dynamics and scalar-quark
 densities themselves. The strength of the intermediate range attraction
 is insensitive to the variation of quark
 masses. The long-range tail of the central CrTPE 
 $NN$ potential is driven by the
 the $\pi N$ sigma term and consistent with the
 effective $\sigma$ meson exchange. In addition, we find a sizable and
 repulsive spin-orbit force which differs from the $LS$ interaction
 generated by the effective $\sigma$ meson in OBE models.

The large-$N_c$ behavior of the CrTPE
 potential was considered and consistency with large-$N_c$ QCD
 expectations was found. Both, the central and $LS$  forces
 satisfy large-$N_c$ QCD counting rules. We have shown that the
 spin-orbit part is $\mathcal{O}(1/N^2_c)$ in strength relative
 to the central force resulting in the ratio $\simeq 1/9$
 suggested by the $1/N_c$ expansion for $N_c=3$. The latter is
 in agreement with our numerical analysis and with
 the Kaplan-Manohar large-$N_c$ power counting. 

Analytical representations for the CrTPE forces in the soft pion
 limit in terms of elementary functions were
 derived and their chiral content was studied. It was shown that the
 latter consists of two terms. The first one is generated by the
 symmetry breaking part of the mesonic Lagrangian and vanishes in
 the chiral limit. The second one, which is non-vanishing in the
 chiral limit, does not contain any pion mass dependence and
 appears as a dominant force which drives all essential features of
 the CrTPE - the strong repulsion at short separations and
 the moderate attraction at intermediate distances. We also find that 
the soft pion limit
provides an accurate representation of the CrTPE and
compares well with direct numerical calculations.

In a forthcoming work the role played by the CrTPE
 in the $NN$ interaction will be addressed again and it
 will be shown that the behavior found here and in
 Ref.~\cite{Oset:2000gn} is an important ingredient to the  $NN$
 interaction and appears to be clearly seen in peripheral $NN$
 phase shifts above the inelastic threshold.

We acknowledge the correspondence with J.R.~Pelaez. 
This work was supported by the
Landesforschungschwerpunkt Baden-W\"urttemberg and the
Bundesministerium f\"ur Bildung und Forschung (06TU201).

\end{document}